# Effect of Couplings weakening and reversing in Ferromagnetic Ising Systems – Rigorous Inequalities

**Moshe Schwartz and Avishay Efrat**

School of Physics and Astronomy

Tel Aviv University,

Ramat Aviv, Tel Aviv 69978, Israel.

**James L. Monroe**

Department of Physics,

Pennsylvania State University,

Beaver Campus,

Monaca, PA 15061-2799, U.S.A.

We consider Ising systems where all the many-spin couplings $J_A$ are positive. We show that the absolute value of all the many-spin correlations does not increase when the value of any of the couplings is reduced, taking any value in the interval $[-J_A, J_A]$. Results of this type are motivated by work in systems such as random field Ising models.

Ferromagnetic Ising systems in the presence of position dependent magnetic fields have attracted a lot of interest over the last three decades, in particular in the context of the random field problem. It turned out that rigorously proven inequalities, such as in [1,2], played an important role in understanding that system and in evaluating various theoretical approaches.

The purpose of the present article is to compare the correlation of spins in the most general ferromagnetic Ising system with the same correlations but in a system in which some or all of the interactions have been reduced. Reduced in such a way that their final values may in fact be negative, although the sum of an interaction in the original system and the corresponding interaction in the reduced system must be non-negative. We prove that the value of a many spin correlation in the original system is greater than or equal to the absolute value of the spin correlation in the reduced system. This result does not follow immediately from previous inequalities such as the second Griffiths, Kelly, Sherman inequality (hereafter GKS) [3,4] because the reduced system can have negative, i.e., anti-ferromagnetic, interactions. It is true that as long as all the couplings in the reduced system remain non-negative then the inequality follows. In fact in this case the correlations are monotonically increasing functions of any coupling. This is not the case if some of the couplings become negative. Indeed, it is not difficult to construct examples in which when some of the couplings become negative the many spin couplings are not monotonic in the coupling strengths.

We will be using the method of "duplicate" variables in our proof. Such a technique has been used to prove a number of previous correlation inequalities. For one dimensional spins, including Ising spins, higher spins such as spin ±1 and 0 systems, and spins allowed to take on a continuous range of values, see the review of correlation inequalities by Sylvester [5].

Consider two systems of Ising spins, each situated on a set of sites $\Omega$, one described by the unprimed Hamiltonian

$$H = -\sum_{A \subseteq \Omega} J_A \sigma_A,\tag{1}$$

the other by the primed Hamiltonian

$$H' = -\sum_{A \subseteq \Omega} J'_A \sigma'_A,\tag{2}$$

and the sum over $A$ is over all subsets of $\Omega$. The notation $\sigma_A$ is used for the product of all spins belonging to $A$,

$$\sigma_A = \prod_{i \in A} \sigma_i.\tag{3}$$

We then have the following theorem.

Theorem: For any subset $A$ of $\Omega$

$$\langle \sigma_A \rangle \geq \left|\langle \sigma_A \rangle'\right|, \qquad \text{if } J_A \geq 0 \text{ and } |J'_A| \leq J_A.\tag{4}$$

where $\langle \ \rangle'$ indicates that the thermal average is for the primed system.

Proof: We need to show that both,

$$\Delta_A^- \equiv \langle \sigma_A \rangle - \langle \sigma_A \rangle' \geq 0 \qquad \underline{\text{and}} \qquad \Delta_A^+ \equiv \langle \sigma_A \rangle + \langle \sigma_A \rangle' \geq 0,\tag{5}$$

or, as may be denoted simply, to show that $\Delta_A^\pm \equiv \langle \sigma_A \rangle \pm \langle \sigma_A \rangle' \geq 0$.

The techniques we will be using in the following are closely related to those reviewed by Sylvester [5]. We start by representing $\Delta_A^\pm$ as

$$\Delta_A^\pm = \frac{\sum_\sigma \sigma_A e^{-\beta H}}{\sum_\sigma e^{-\beta H}} \pm \frac{\sum_{\sigma'} \sigma'_A e^{-\beta H'}}{\sum_{\sigma'} e^{-\beta H'}} = \frac{\sum_\sigma \sum_{\sigma'} (\sigma_A \pm \sigma'_A) e^{-\beta H} e^{-\beta H'}}{\sum_\sigma \sum_{\sigma'} e^{-\beta H} e^{-\beta H'}}.\tag{6}$$

The denominator, as a sum of positive terms, is obviously positive, and therefore, we need to show that

$$\tilde{\Delta}_A^{\pm} \equiv \sum_{\sigma}\sum_{\sigma'}(\sigma_A \pm \sigma'_A)e^{-\beta H}e^{-\beta H'} \geq 0. \tag{7}$$

In order to do so, we define for each site $i \in \Omega$

$$t_i \equiv \frac{1}{2}(\sigma_i + \sigma'_i) \quad \text{and} \quad q_i \equiv \frac{1}{2}(\sigma_i - \sigma'_i), \tag{8}$$

so that

$$\sigma_i = t_i + q_i \quad \text{and} \quad \sigma'_i = t_i - q_i. \tag{9}$$

Substituting $t$'s and $q$'s, for all $\sigma$'s and $\sigma'$'s, in Eq. (7), we have

$$\tilde{\Delta}_A^{\pm} = \sum_{\{\sigma\}}\sum_{\{\sigma'\}}\left\{\prod_{i\in A}(t_i + q_i) \pm \prod_{i\in A}(t_i - q_i)\right\} \times \\ \times \prod_{B\subseteq\Omega}\exp\left[\beta J_B \prod_{j\in B}(t_j + q_j) + \beta J'_B \prod_{j\in B}(t_j - q_j)\right]. \tag{10}$$

Consider first the curly brackets in the equation above. For any choice of the sign, it is clear after expanding the products that any term with a negative prefactor, coming from the second product, is canceled by a matching term with a positive prefactor, coming from the expansion of the first product. The result is that only multiples of $t$'s and $q$'s with positive prefactors are left. Consider next a single exponent in the product over $B$. Imagine then expanding the products in the exponents as above. When expanding the second product in the exponent, some of the products of $t$'s and $q$'s have a prefactor that is $\beta J'_B$ and some have a negative prefactor that is $-\beta J'_B$. As a result, the whole exponent is made up of a sum of products of $t$'s and $q$'s, some having a prefactor that is $\beta(J_B + J'_B)$ and some having a prefactor that is $\beta(J_B - J'_B)$. Because of the restrictions in (4), both, $J_B + J'_B$ and $J_B - J'_B$, are positive. So, in the exponent we have the sum of products of $t$'s and $q$'s with positive prefactors. Next we expand the exponential and then take the product of exponentials. The result is still a sum of products of $t$'s and $q$'s with positive prefactors.

We show next that, when summing over configurations of the two spin systems, each of these terms gives either zero or a positive contribution. In order to do so, we collect for each of these terms all the *t*'s and *q*'s by site. Thus each site is represented in each of these terms by a factor of the form

$$\Sigma_i \equiv \sum_{\sigma_i=\pm1}\sum_{\sigma'_i=\pm1} t_i^\alpha q_i^\gamma = \sum_{\sigma_i=\pm1}\sum_{\sigma'_i=\pm1} (\sigma_i+\sigma'_i)^\alpha (\sigma_i-\sigma'_i)^\gamma .\qquad(11)$$

As our concluding step, we show now the r.h.s. of Eq. (11) to be non-negative. Calculating it explicitly we have

$$\Sigma_i = \begin{cases} 0 & \alpha\neq 0 \text{ and } \gamma\neq 0 \\ \sum_{\sigma_i=\pm1}\sum_{\sigma'_i=\pm1}(\sigma_i-\sigma'_i)^\gamma = 2^\gamma + (-2)^\gamma \geq 0 & \alpha=0 \text{ and } \gamma\neq 0 \\ \sum_{\sigma_i=\pm1}\sum_{\sigma'_i=\pm1}(\sigma_i+\sigma'_i)^\alpha = 2^\alpha + (-2)^\alpha \geq 0 & \alpha\neq 0 \text{ and } \gamma= 0 \end{cases} .\qquad(12)$$

This completes the proof.

Comparing the l.h.s. of Eq. (11) with a very similar expression given by Sylvester (l.h.s. of Eq. (22) in [5]), we believe that the theorem proven here can be extended to the much larger class of spin systems, considered by Sylvester.

The first two authors have an independent proof using a ghost spin approach similar to that in [3].

**Acknowledgement:** This work was supported in part by a grant from the Israeli Foundation for basic research.